\documentclass[11pt]{article}
\usepackage{moriond,epsfig}

\def\Journal#1#2#3#4{{#1} {\bf #2}, #3 (#4)}

\def\NIMA{{\em Nucl. Instrum. Methods} A}

\def\MST{\em Meas. Sci. Technol}
\def\JOP{\em J. Phys. G: Nucl. Part. Phys.}
\def\ASP{\em Astropart. Phys.}

\def\be{\begin{equation}}
\def\ee{\end{equation}}
\def\bea{\begin{eqnarray}}
\def\eea{\end{eqnarray}}

\begin{document}
\vspace*{4cm}
\title{TACTIC and MACE gamma-ray telescopes}

\author{ K.K. Yadav (for the HIGRO collaboration)}

\address{Astrophysical Sciences Division, Bhabha Atomic Research Centre,\\
Trombay, Mumbai, India - 400 085}

\maketitle\abstracts{
The TACTIC gamma-ray telescope, equipped with a tracking light collector of $\sim$9.5m$^2$ area and a 349-pixel imaging camera has been in operation at Mount Abu in Western India since 2001. Having a sensitivity of detecting the Crab Nebula above 1.2 TeV at 5.0$\sigma$ significance level in 25h of observations, this telescope has detected gamma-ray emissions from Mrk501 and Mrk421 and is presently being deployed for monitoring of AGNs. As a new Indian initiative in $\gamma$-ray astronomy we are setting up the 21-m diameter MACE $\gamma$-ray telescope at the high altitude (4200m asl) astronomical site at Hanle in North India. This telescope will deploy a 1408-pixels integrated camera at its focal plane. Designed to operate at a trigger threshold of $\sim$30 GeV, this telescope is expected to be operational in 2011. Some of the salient features of the TACTIC telescope along with the results of its recent observations and the design details of the MACE telescope are presented in this paper.
}

\section{TACTIC telescope}
The TACTIC (TeV Atmospheric Cherenkov Telescope with Imaging Camera) $\gamma$-ray telescope  located at Mt. Abu (24.6$^\circ$ N, 72.7$^\circ$ E, 1300m asl), is being used to study potential TeV  $\gamma$-ray sources. The telescope deploys a F/1 type tracking light collector of $\sim$9.5 m$^2$ area, made up of 34$\times$0.6 m diameter, front-coated spherical glass facets which have been prealigned to produce an on-axis spot of $\sim$ 0.3$^\circ$ diameter at the focal plane. The telescope uses a 349-pixel photomultiplier tube (ETL 9083UVB) -based imaging camera  with a uniform pixel resolution of $\sim$0.3$^\circ$ and a  field-of-view of $\sim$6$^\circ\times$6$^\circ$  to record images of atmospheric Cherenkov events. The innermost 121 pixels (11$\times$11 matrix) are used for  generating the event trigger, based on the NNP (Nearest Neighbour Pairs)/3NCT (Nearest Neighbour Non-Collinear Triplets) topological logic ~\cite{sr}, by demanding a signal $\geq$ 25/8 pe for the 2/3 pixels which participate in the trigger generation. Whenever the single channel rate of any two  or  more pixels in the trigger region goes outside the preset operational band, it is automatically restored to within the prescribed range by appropriately adjusting the high voltage of the pixels ~\cite{nb}. The resulting change in the photomultiplier (PMT) gain is monitored by repeatedly flashing a blue LED, placed at a distance of $\sim$1.5m from the camera. The advantages of using such a scheme are that in addition to  providing control over chance coincidence triggers, it also ensures  safe operation of PMTs with typical anode currents of $\leq$ 3 $\mu$A. The back-end signal processing hardware of the telescope is based on medium channel density  NIM and CAMAC  modules developed inhouse. The data acquisition and control system of the telescope~\cite{ky1} has been designed around a network of PCs running the QNX (version 4.25) real-time operating system. The triggered events are digitized by CAMAC based 12-bit Charge to Digital Converters (CDC) which have a full scale range of 600 pC. The telescope has a pointing and  tracking accuracy of better than $\pm$3 arc-minutes. The tracking accuracy  is checked  on a regular basis  with so called "point runs", where a bright  star whose declination is close to that of the candidate $\gamma$-ray source is  tracked continuously  for about 5 hours.  The  point run calibration data  (corrected zenith and azimuth angle  of the telescope  when the star image is centered)  are  then incorporated in the telescope drive system software so that  appropriate corrections can be  applied directly  in real time  while tracking  a candidate $\gamma$-ray source~\cite{rk}. 
 
\begin{table}[t]
\caption{Observations on gamma-ray sources with TACTIC telescope}
\vspace{0.4cm}
\begin{center}
\begin{tabular}{|c|c|c|c|c|}
\hline
Sr. & Source & Observation period & Observation(h) & Significance/UL\\
\hline
1 & Crab Nebula  & Dec 2003 - Feb 2004& 104.28&10.30$\sigma$ \\
\cline{3-5}
&  & Nov 2005 - Feb 2006& 101.04 &9.40$\sigma$ \\
\cline{3-5}
& & Nov 2007 - Mar 2008 & 105.15&11.05$\sigma$ \\
\hline
2 & Mrk421 & Dec 2005 - Apr 2006 & 201.72& 11.5$\sigma$\\
\cline{3-5}
&  & Jan 2007 - Mar 2007& 83.5 &$\leq 0.92\times10^{-12}$ $ ph$ $ cm^{-2} s^{-1}$ \\
\cline{3-5}
&  & Jan 2008 - May 2008& 149.70 &9.60$\sigma$ \\
\hline
3 & Mrk501 & Mar 2005 - May 2005 & 46.00 &$\leq 4.62\times10^{-12}$ $ ph$ $ cm^{-2} s^{-1}$\\
\cline{3-5}
&  & Feb 2006 - May 2006 & 66.80 &7.5$\sigma$ \\
\hline
4 & 1ES2344+514 &Oct 2004 - Dec 2005 &60.15 & $\leq 3.84\times10^{-12}$ $ ph$ $ cm^{-2} s^{-1}$\\
\hline
5 & H1426+428 &Mar 2004 - Jun 2007 &165.70 & $\leq 1.18\times10^{-12}$ $ ph$ $ cm^{-2} s^{-1}$\\
\hline
\end{tabular}
\end{center}
\end{table}
\subsection{Recent TACTIC results}
In order to evaluate the performance of the TACTIC telescope the Crab Nebula ``standard candle'' has been observed repeatedly since 2001. Operating at a $\gamma$-ray  threshold energy of $\sim$1.2 TeV, the telescope records a  cosmic ray event rate of $\sim$2.0 Hz at a typical zenith angle of 15$^\circ$. The telescope has a 5$\sigma$ sensitivity of detecting Crab Nebula in 25 hours of observation time. The consistent detection of a steady signal from the Crab Nebula along with excellent matching of its energy spectrum with that obtained by other groups, reassures that the performance of the TACTIC telescope is quite stable and reliable. The telescope has detected strong $\gamma$-ray signals from two active galactic nuclei (AGN) Mrk501 (2006 observations)~\cite{sv2} and Mrk421 ( 2005-06 observations)~\cite{ky2} while other two AGNs 1ES2344+514~\cite{sv1} and H1426+428 observed during 2004-05 and 2004-07 respectevely have been found to be in the quiescent state. Some of the recent results obtained on various candidate $\gamma$-ray sources are listed in Table 1. We believe that there is a considerable scope for the TACTIC telescope to monitor TeV $\gamma$-ray emission from other AGNs on a long-term basis.

\section{MACE telescope }
Exploring the $\gamma$-ray sky in the energy range $\geq10 GeV$ with low energy threshold ground based atmospheric Cherenkov telescopes is expected to lead to a potentially rich harvest of astrophysical discoveries, as has been already demonstrated by the HESS and MAGIC telescopes at $\gamma$-ray energies $\geq100 GeV$.
The low threshold energy can be attained by increasing the light collector area of the telescopes and installing them at higher altitudes where the photon density of the atmospheric Cherenkov events is higher~\cite{ar}. As a new Indian initiative in gamma-ray astronomy, the Himalayan Gamma Ray Observatory (HIGRO) is being set up at Hanle (32.8$^\circ$ N, 78.9$^\circ$ E, 4200m asl) in the Ladakh region of North India. The site offers an average of about 260 uniformly distributed spectroscopic nights per year which is a major advantage in terms of sky coverage for source observations. Located closer to the shower maximum the Cherenkov photon density at Hanle is substantially high as compared to the sea level~\cite{co}. The higher photon density along with the low background light level at this site helps in lowering the energy threshold of the Cherenkov telescope being setup there.
\par 
The MACE (Major Atmospheric Cherenkov Experiment) telescope with high resolution imaging camera is designed to operate in the sub-TeV energy range as part of the HIGRO collaboration. As depicted in Figure 1 the altitude-azimuth mounted telescope will deploy a 21-m diameter parabolic light collector made of 356 panels of 984 mm $\times$ 984 mm size with each panel consisting of 4 spherical mirror facets of 488 mm $\times$ 488 mm size. Each facet is diamond turned to a mirror finish yielding a reflectivity of $\geq$ 85$\%$ in the visible band. The telescope will use the graded focal length (increases towards the periphery) mirrors in order to reduce the D$_{80}$ spot size (defined as the diameter of the circle within which 80$\%$ of the reflected rays lie) of the light collector to $\sim$15 mm for on-axis incidence. Each mirror panel will be equipped with motorized orientation controllers for aligning them to form a single parabolic light collector.
\par
The focal plane instrumentation will have a photomultiplier tube based imaging camera covering a field of view of 4$^\circ\times$4$^\circ$. The imaging camera will comprise of 1408 pixels arranged in a square matrix with uniform pixel resolution of 0.1$^\circ$. The inner 576 pixels with field of view of 2.4$^\circ\times$2.4$^\circ$  will be used for generating the event trigger. The PMTs will be provided with acrylic front-aluminized light cones for enhancing the light collection efficiency of the camera. The signal processing instrumentation will also be housed within the camera and the acquired data will be sent to the control room over the computer network for processing and archiving. Detailed Monte Carlo simulation studies have been carried out using CORSIKA~\cite{he}code and the results suggest that using a pixel threshold of $\geq$4pe and a 4 nearest neighbour pixel trigger, gamma-ray energy threshold of $\sim$30 GeV is achievable by the MACE telescope. Figure 2 shows the differential trigger rates of $\gamma$-rays for the two different types of spectra. The energy thresholds are determined to be $44 GeV$ for the Crab spectrum and $31 GeV$ for the pure power law spectrum with a diffential index of 2.59 for the above mentioned configuration.

\begin{figure}
\centering
\includegraphics[width=0.45\textwidth]{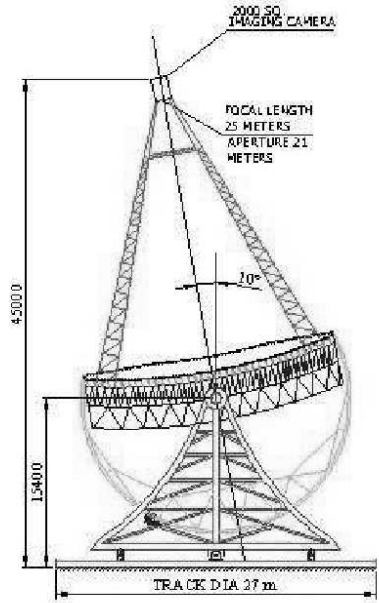}
\caption{21-m diameter MACE telescope}
\label{Fig:Mace}
\end{figure}

\begin{figure}
\centering
\includegraphics[width=0.45\textwidth,angle=-90]{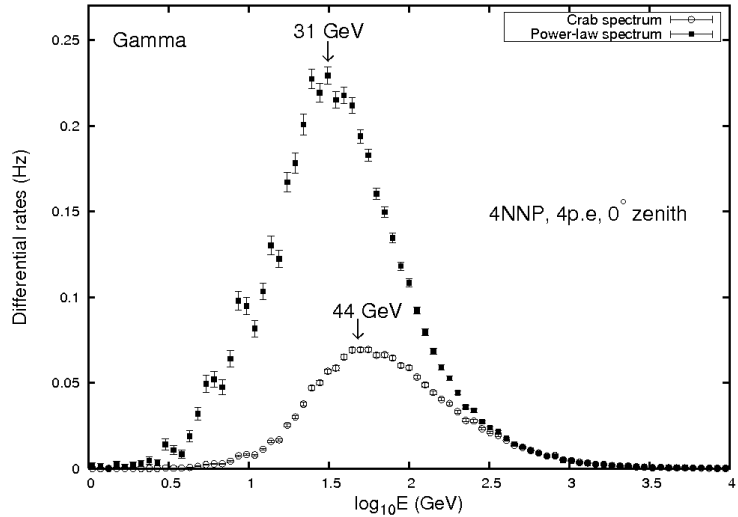}
\caption{Gamma-ray differential rates for the two types of primary spectra calculated for the 4 nearest neighbour pixel, 4pe trigger configuration. For Crab, it gives an energy threshold of $\sim(44\pm2) GeV$ and for the power law the threshold energy is $\sim(31\pm2) GeV$.}
\label{Fig:Mace}
\end{figure}
\subsection{Status of MACE telescope}
The detailed engineering and structural design of the MACE telescope has been completed. Fabrication of the mechanical structure has started and the telescope is likely to be installed at Hanle by 2011.

\section*{Acknowledgments}
We would like to thank the organizers of the ``44th Rencontres de Moriond'' for providing financial support to attend the conference. We would also like to thank colleagues of the Astrophysical Sciences Division for their contribution towards the operation of the TACTIC telescope and the design of the MACE telescope.

\end{document}